\documentstyle[12pt]{article}

\def \be{\begin{equation}}
\def \ee{\end{equation}}
\def \bea{\begin{eqnarray}}
\def \eea{\end{eqnarray}}

\begin{document}
\begin{center}
{\Large{\bf{The Exact N-Point Generating Function in Polyakov - Burgers
Turbulence
}}}

\vskip 2cm
{\large{M.R.Rahimi Tabar, S.Rouhani and B.Davoudi}}
\vskip 1cm
{\it{Department of Physics, Sharif University of Technology\\Tehran P.O.Box:
11365-9161, Iran.\\Institue for Studies in Theoretical Physics and Mathematics
\\ Tehran P.O.Box: 19395-5746, Iran.}}
\end{center}
\vskip 1.5cm
\begin{abstract}
We find the exact N-point generating function in Polyakov's approach
to Burgers turbulence.
\end{abstract}
\newpage
{\bf 1- Introduction}

A theoretical understanding of turbulence has eluded physicists for a long
time.
Recently Polyakov [1] has offerd a field theoretic method for  deriving of the
probability distribution or density of states in (1+1)-dimensional turbulent
systems. Polyakov
formulates a new method for analyzing the inertial range correlation
functions based on the two important ingredients in field theory and
statistical physics namely the operator product expansion (OPE) and anomalies.
Despite existence of many field theoretic approaches to turbulence [2,3,4],
it appears that this new approach is more promising.
Polyakov argues that in the limit of high Reynold's number because of
existence of singularities at coinciding point, dissipation remains finite and
all sublleading terms give vanishing contributions in the inertial range.
By develop the OPE one finds the leading singularities and
can show that this approach is self-consistent.
Here we consider  Polyakov`s approach [1] to the Burger turbulence when
the gradient of pressure is negligible and solve the N-point master
equation, calculating the  N-point generating function. Our
result also apply to the Kardar-Parisi-Zhang (KPZ) equation in
(1+1)-dimensions, crystal growth [5], the nonlinear dynamics of a moving line
[6], galaxy formations [7], dissipative transport [9], dynamics of a
Sine-Gordan
chain [10], behavior of magnetic flux line in superconductor [11], and spin
glasses [12].
\\
\\
{\bf 2-  N-Point Generating Functions} \\
The Burgers equation has following form
\be
u_t + uu_x = \nu u_{xx} + f(x,t)
\ee
where $u$ is the velocity field, and $\nu$ is the viscosity and $f(x,t)$ is the
Gaussian random force with the following correlation:
\be
<f(x,t) f(x^{'},t^{'})> = k (x-x^{'}) \delta (t-t^{'})
\ee
The transformation, $u(x,t) = -\lambda \partial_x h(x,t)$ maps eq.(1) to the
well-known KPZ equation [5],
\be
\partial_t h = \nu \partial_{xx} h + {\lambda \over 2}[\partial_x h]^2 +
f(x,t)
\ee
It is noted that the parameter $\lambda$ that appears in the above
transformation is not
renormalized under any renormalization procedure [8].
Following Polyakov [1] consider the following generating functional
\be Z_N(\lambda_1, \lambda_2,\ldots \lambda_N, x_1,\ldots x_N) = <\exp (
\sum^N_{j=1} \lambda_j u(x_j,t))>
\ee
Noting that the random force $f(x,t)$ has a Gaussian distribution
$Z_N$ satisfies a closed differential
equation provided that the viscosity  $\nu$ the tends to zero:
\be
\dot{Z_N} + \sum \lambda_j {\partial \over {\partial \lambda_j}}({1\over
{\lambda_j}}{\partial Z_N \over{\partial x_j}}) = \sum k(x_i-x_j) \lambda_i
\lambda_j Z_N +D_N
\ee
where $D_N$ is :
\be
D_N = \nu \sum \lambda_j <u^{''}(x_j,t) \exp \sum \lambda_k u(x_k,t)>
\ee
To reach the inertial range we must, how ever, keep
$\nu$ infinitesimal but non-zero. Polyakov argues that the anomaly mechanism
implies that infinitesimal viscosity produces a finite effect. To compute
this effect Polakov makes the F-conjecture, which is the existence of an
operator
product expansion or the fusion rules. The fusion rule is the statement
concerning the behaviour of correlation functions, when some subset of
points are put close together.

Let us use the following notation;
\be
Z(\lambda_1, \lambda_2,\ldots , x_1,\ldots x_N)= <e_{\lambda_1}(x_1)\ldots
e_{\lambda_N}(x_N)>
\ee
then Polyakov`s F-conjecture is that in this case the OPE has the following
form,
\be
e_{\lambda_1}(x+y/2) e_{\lambda_2}(x-y/2) = A(\lambda_1, \lambda_2, y)
e_{\lambda_1+\lambda_2}(x)+B(\lambda_1, \lambda_2, y){\partial \over \partial
x}
e_{\lambda_1+\lambda_2}+o(y^2)
\ee
This implies that $Z_N$ fuses
into functions $Z_{N-1}$ as we fuse a couple of points together. The
F-conjecture
allows us to evaluate the following anomaly operator (i.e. the $D_N$-term in
eq.(5)),
\be
a_\lambda(x)= \lim _{\nu\rightarrow 0}\nu (\lambda u^{``}(x) \exp (\lambda
u(x))
\ee
which can be written as:
\be
a_\lambda(x)= \lim _{\xi,y,\nu\rightarrow 0}\lambda \nu {\partial^3\over
{\partial \xi \partial y^2}} e_\xi (x+y) e_\lambda(x)
\ee
As discussed in [1] the only possible Galilean invarant expression is:
\be
a_\lambda(x) = a(\lambda) e_\lambda(x) + \tilde{\beta}(\lambda){\partial
\over \partial x}e_\lambda(x)
\ee
Therefore in steady state the master equation takes the following form,
\bea
\sum ({\partial \over \partial x_j} - \beta(\lambda_j)){\partial \over
\partial x_j} Z_N -\sum k(x_i -x_j) \lambda_i \lambda_j Z_N &=& \sum
a(\lambda_j) Z_N \cr
\beta(\lambda) &=& \tilde{\beta}(\lambda)+{1 \over \lambda}
\eea

Polyakov has found the explicit form of $Z_2$ in the case that $k(x_i - x_j )=
K(0)(1-(x_i - x_j)^2/{l^2})$, and the desity of states as;

\be
W(u,y) = \int_{c-i\infty}^{c+i\infty} {d\mu \over 2\pi i} e^{-\mu u}
\phi(\mu y)
\ee
where
$$ \phi(\mu y) = e^{2/3(\mu y)^{3/2}}$$
and $$\mu=2 (\lambda_1 -\lambda_2), y=x_1-x_2.$$
It can be easly shown that with the following definition  of variables
Polyakov`s
master equation with the scaling conjecture[1] is:

\be
\{{\partial^2 \over {\partial \mu_2 \partial y_2}}+ {\partial^2 \over
{\partial \mu_3 \partial y_3}}-(y_2 \mu_2 +y_3 \mu_3)^2\} f_3 =0
\ee
where
$$f_3 = (\lambda_1 \lambda_2 \lambda_3)^{-b}Z_3$$
\bea
y_1 &=& {x_1+x_2+x_3 \over 3} \hskip 0.5cm y_2= x_1-{x_2+x_3 \over 2}
\hskip 0.5cm y_3 =x_2-x_3 \cr
\mu_1 &=& {\lambda_1+\lambda_2+\lambda_3 \over 3} \hskip 0.5cm
\mu_2= {3 \over 2}(\lambda_1-{\lambda_2+\lambda_3 \over 2})
\hskip 0.5cm \mu_3 =2(\lambda_2-\lambda_3)
\eea
Now we set $f$ as;
\be
f_3 = \mu_2^{S_2} \mu_3^{S_3}g_3(\mu_2 y_2,\mu_3 y_3)
\ee
inserting in eq.(15) results in
\be
g_3(\mu_2 y_2, \mu_3, y_3) = e^{2/3(\mu_2 y_2+ \mu_3 y_3)^{3/2}}
\ee
and $$ S_2 = S_3 =-5/4$$
Now if we use the following transformation;
\bea
y_1&=&{{x_1 +x_2 +x_3 +\ldots x_N}\over N}\cr
y_2&=&x_1 -{{x_2 +x_3 +\ldots x_N}\over N-1}\cr
y_3&=&x_2 -{{x_3 +x_4 +\ldots x_N}\over N-2}\cr
and \hskip 1cm y_N&=& x_{N-1}-x_N
\eea
and
\bea
\mu_1 &=&{{\lambda_1 + \lambda_2 +\ldots +\lambda_N}\over N}\cr
\mu_2 &=&{N\over N-1}[\lambda_1-{{\lambda_2 + \lambda_3 +\ldots +\lambda_N}
\over {N-1}}]\cr
\mu_3 &=&{N-1 \over{N-2}}[\lambda_2-{{\lambda_3 + \lambda_4 +\ldots +\lambda_N
}\over {N-2}}]\cr
and \hskip 1cm \mu_N&=& 2(\lambda_{N-1} - \lambda_N)
\eea
we get the following partial differential equation for $f_N$;
\be
\{{\partial^2 \over {\partial y_2 \partial \mu_2}} + \ldots +{\partial^2
\over {\partial y_N \partial \mu_N}}\} f_N - (y_2 \mu_2+\ldots +y_N \mu_N)^2
f_N = 0
\ee
which is solved by:
\be
f_N = (\mu_2 \mu_3 \ldots \mu_N)^{-[{2N-1 \over {2(N-1)}}]} \exp^{2/3(
\mu_2 y_2 + \ldots + \mu_N y_N)^{3/2}}
\ee
In principle the parameter $b$ in eq.(15) can be evaluated by
means of the exponents of $(\mu_2 \ldots \mu_N)$ term in eq.(21) which for
$Z_2$
it turns out that $b=3/4$. Our exact result for $Z_N$ allows us to determaine
the
OPE coefficients in eq.(8), and calculate $Z_N$ in the case where the random
force
is conservative, which is important for the KPZ-equation, work in this
direction is
on the way.
\\
\\
{\bf Acknowledgements:} We wish to thank Vahid Karimipour for valuable
discussions.

\end{document}